\documentstyle [12pt]{ARTICLE}
\begin{document}

\def\be{\begin{equation}}
\def\ee{\end{equation}}
\def\sd{\strut\displaystyle}
\begin{flushright}
UAB-FT-345/94\\
October 10th-1994\\
\end{flushright}
\vskip 1.2cm
\begin{center}
{\bf PAST-FUTURE INTERFERENCE IN $\Phi$-DECAYS }\vskip .5 cm
{\bf INTO ENTANGLED KAON PAIRS }\vskip 1.5 cm
Bernat Ancochea and Albert Bramon \footnote{16414::BRAMON}\\
\vskip .1 cm
Grup de F\'\i sica Te\`orica, Universitat Aut\`onoma de Barcelona,\\
 08193 Bellaterra (Barcelona), Spain

\vskip  6 cm
{\bf Abstract}
\end{center}
\par
Quantum interference effects between past and future events
(neutral kaon transitions, including CP violating decays) are
discussed for entangled kaon pairs produced through
$\phi \to K^0 \bar{K^0}$ in $\phi$-factories.
Contrasting with the exciting conclusions
of other recent analyses,
we predict the inexistence of such an observable effect.

\pagestyle{plane}
\setcounter{page}{1}
\newpage

{\bf 1. Introduction}
\par
The possibility of faster-than-ligth communication and the related, and
even more intriguing, prospective that future events could have an
influence over present ones, have been discussed several times in
physics. For a manifestation, if any, of such a  highly
speculative violation of causality, the concurrence of several very
special and counter-intuitive circumstances should (at least) be
required. The experimental success and the
well-known non-locality of quantum mechanics -- as
clearly manifested in Einstein-Podolsky-Rosen (EPR) experiences -- is
sometimes considered as a rather promising one. A second,
favourable circumstance could be related to the non-invariance of
time-reversal, as observed (and theoretically well described) in
neutral-kaon decays. Both of these two unusual
circumstances concur when dealing with $\phi$-meson transitions
into (EPR) entangled states containing two neutral-kaons which
subsequently decay through time-reversal- or CP-violating channels.
A high-luminosity $\phi$-factory
-- such as DA$\Phi$NE, under construction in Frascati
\cite{dafne} -- is therefore an exceptionally appropriate machine
for this kind of studies. In this $\phi$-factory context,
Datta, Home and Raychaudhuri \cite{datta}, \cite{datta2}
proposed some time ago the
intriguing possibility of detecting such a curious non-local
and faster-than-light propagating
effect on the statistical (i.e., observable) level.
Their result was criticized by several
authors \cite{several} on general grounds and by Ghirardi et al.
\cite{ghirardi} in a more detailed way, but the controversy appears to
be still open \cite{datta2}.
\par
More recently, Srivastava and Widom \cite{yogi}
have presented a detailed and interesting discussion
on a new experiment (somehow related to that considered
by Datta et al \cite{datta}) leading again to an intriguingly
positive result. A relevant feature of this new proposal is that it
fits perfectly well with the DA$\Phi$NE configuration and,
on the theoretical side, only well tested and extensively studied
equations (specially by workers preparing DA$\Phi$NE experiments)
are used. Indeed, one starts considering $\phi$-decays at $t = 0$
into entangled neutral-kaon pairs,
\be
\mid \Phi (t=0) \rangle = \frac{1}{\sqrt2} \left[ \mid K^0 \rangle
\mid \bar {K^0} \rangle - \mid \bar{K^0} \rangle \mid K^0 \rangle \right],
\ee
where the relative minus sign comes from the well-known negative
charge-conjugation of the initial $\phi$. The kaons fly apart in
opposite directions along a given axis, thus defining a left
and a right beam. Kaons along the left beam travel in free-space
from the source, at $t=0$, up to a nearby detector of
neutral-kaon decays into a given channel (say, c-channel) placed
at a "time (of flight) distance" $t$; the number of c-channel decays
per unit of time, $dP_c/dt$, at $t$ is measured there.
Kaons in the right beam fly freely from the common source, $t=0$,
up to the edge of a distant absorber, placed at a "time-distance"
$T > t$. The essential claim in ref. \cite{yogi}
(quite in line with that in refs. \cite{datta}, \cite{datta2})
is that measurements performed on the left
beam at a "time-distance" $t$ can be statistically modified
by the presence or not of the kaon absorber
located at a larger "time-distance" $T$, $T > t$, on the other beam.
In other words, that $dP_c(t,T)/dt$ depends not only
on $t$ (as it obviously does), but $also$ on $T$, $i.e.$, on
"where" on the other side (future) absorption starts
(or immediatelly takes places if an "ideal" absorber is used).
The analysis by Srivastava et al. is somehow simplified by
the use of an "ideal" (or drastic) absorber and by neglecting
(irrelevant) kaon decay phases. But, most interestingly, the
analysis holds for any space-time separation of events, including
c-channel decays at $t$ belonging to the absolute past of
the respective absorption events at $T$. In this sense, one is dealing
with the Einstein-Tolman-Podolsky effect \cite {etp} rather than with
the more familiar EPR-effect discussed by Datta et al. \cite{datta}.
The unexpectedly predicted T-dependence is then enhanced by performing a
time-$t$
integration from $t_i$ to $t_f$ under specific $T$-independent
conditions. This final step, which is not
essential for our main discussion (i.e., if $dP_c/dt$
is $T$-dependent or not), enhances the effect up to about $10^{-4}$,
thus allowing for possible detection at DA$\Phi$NE.
\par
Our purpose in this note is to present a detailed and accurate
rediscussion of the whole situation taking all the phases
into account and substituting the ideal absorber at $T$ by a
more general and realistic
one absorbing homogenously from $T$ to infinity. To this aim, the
introduction of three different basis for the neutral-kaon system is
required. These standard basis are discussed in the next paragraph
along the lines of ref. \cite{kabir}. Then we proceed to explicitly
calculate $dP_c/dt$ for a general, homogenous absorber
and a final section is devoted to a discussion including the
particular case of "ideal" absorption considered in \cite{yogi}.
\vskip 1cm
{\bf 2. Basis for neutral kaons}
\par
$i$) The first, "strong-interaction" basis contains the two eigenstates
$\mid {K^0} \rangle$ and $\mid \bar{K^0} \rangle$, with strangeness $S=+1$ and
$-1$, and, contrasting with the next two basis,
it is the only orthogonal one,
$\langle K^0 \mid \bar{K^0} \rangle = 0$. This is the suitable basis to discuss
$S$-conserving strong interactions, but not to consider simple
neutral kaon evolution in free space or inside matter; in both of these
later cases, conversions or $K^0-\bar{K^0}$ oscillations take place.
\par
$ii$) The "free-space" basis is defined by the $K$-short and $K$-long
states
\begin{eqnarray}
\mid K_S \rangle &=& (1 + \mid r \mid ^2)^{-1/2} \left[ \mid {K^0} \rangle
+ r \mid \bar{K^0} \rangle  \right] \nonumber
  \\
\mid K_L \rangle &=& (1 + \mid r \mid ^2)^{-1/2}  \left[ \mid {K^0} \rangle
- r \mid \bar{K^0} \rangle  \right],
\end{eqnarray}
where $r \equiv (1-\epsilon)/(1+\epsilon)$ and $\epsilon$ is the usual
CP-violation parameter. $\mid K_S \rangle$ and $\mid K_L \rangle$ are the
normalized eigenvectors of the effective weak hamiltonian
\be
H = \pmatrix{ \lambda_+ & \lambda_-/ r \cr
              \lambda_- r  & \lambda_+ \cr },
\ee
with eigenvalues
\begin{eqnarray}
\lambda_S &=& \lambda_+ + \lambda_- = m_S - (i/2) \Gamma_S \nonumber \\
\lambda_L &=& \lambda_+ - \lambda_- = m_L - (i/2) \Gamma_L,
\end{eqnarray}
where $m_{S,L}$ and $\Gamma_{S,L}$ are the physical
masses and decay widths of $K_{S,L}$, and
PCT-invariance requires the diagonal elements of $H$ be equal.
This is the appropriate basis to discuss propagation in free space.
Indeed, $K_S$ and $K_L$ states do not convert into each other and
show simple exponential decay in time
\be
\mid K_S(t)\rangle = e^{-i\lambda_S t} \mid K_S \rangle,\ \ \ \
\mid K_L(t)\rangle = e^{-i\lambda_L t} \mid K_L \rangle
\ee
The non-orthogonality of this basis is given by
\be
\delta \equiv \langle K_L \mid K_S \rangle = \langle K_S \mid K_L \rangle =
\frac{1-\mid r \mid^2}{1+ \mid r \mid^2} =
\frac{\epsilon + \epsilon^*}{1 + \mid \epsilon \mid^2}
\ee
and simple unitarity arguments lead to the so-called Bell-Steinberger
relation \cite{kabir}
\be
\delta = \left( \sum_f A^*_{K_S \to f} A_{K_L \to f} \right)
/ (\Gamma + i \Delta m)
\ee
with $\Gamma + i \Delta m \equiv i ( \lambda_L - \lambda_S^*) =
(\Gamma_L + \Gamma_S)/2 + i (m_L - m_S)$ and the
sum extending over all possible decay amplitudes $A_{K_{L,S} \to f}$.
\par
$iii$) The "inside-matter" basis is given by the $K'_S, \ K'_L$
normalized states
\begin{eqnarray}
\mid K'_S \rangle &=& (1 +\  \mid r \bar{\rho} \mid ^2)^{-1/2}  \left[
\mid {K^0} \rangle + r  \bar{\rho}\ \mid \bar{K^0} \rangle  \right] \nonumber
  \\
\mid K'_L \rangle &=& (1 + \mid  r / \bar{\rho} \mid ^2)^{-1/2}  \left[
 \mid {K^0} \rangle - (r / \bar{\rho}) \mid \bar{K^0} \rangle  \right],
\end{eqnarray}
where we have introduced the regeneration parameter $\rho$
\cite{kabir}, and
$\bar{\rho}$,
\be
\bar{\rho} \equiv \sqrt{1+ 4 \rho^2} + 2 \rho,\ \ \ \ \
\rho \equiv \frac{\pi \nu}{m_K} \frac{f - \bar f}{\lambda_S - \lambda_L}
{}.
\ee
The $K'_S, \ K'_L$ states diagonalize the effective weak plus strong
interaction hamiltonian
\be
H' =  H - \frac{2 \pi \nu}{m_K} \pmatrix{ f & 0 \cr
                                          0 &  \bar f \cr},
\ee
where $f$ and $\bar f$ are the forward scattering amplitudes for $K^0$
and $\bar{K^0}$ on nucleons and $\nu$ is the nucleonic density in the
homogeneous absorber. The corresponding eigenvalues are
\begin{eqnarray}
\lambda'_S &=& \lambda_+ - \frac{\pi \nu}{m_K} (f + \bar f)
                         + \lambda_- \sqrt{1 + 4 \rho^2}, \nonumber \\
\lambda'_L &=& \lambda_+ - \frac{\pi \nu}{m_K} (f + \bar f)
                         - \lambda_- \sqrt{1 + 4 \rho^2}.
\end{eqnarray}
The $K'_S, \ K'_L$ eigenstates are appropriate to discuss propagation
inside a homogeneous medium  showing no-reconversion into each other
and simple exponential extinction in time
\be
\mid K'_S(t)\rangle = e^{-i\lambda'_S t} \mid K'_S \rangle,\ \ \ \
\mid K'_L(t)\rangle = e^{-i\lambda'_L t} \mid K'_L \rangle
\ee
due to both weak decays and strong interaction scattering out of
the beam. The non-orthogonality now reads
\begin{eqnarray}
\delta'& \equiv & \langle K'_L \mid K'_S \rangle \  = \  \langle K'_S \mid K'_L
\rangle^*
\nonumber \\
&=& \left (1+ \mid r \bar{\rho} \mid^2 \right)^{-1/2}
\left (1+ \mid r / \bar{\rho}\mid^2 \right)^{-1/2}
\left (1- \mid r \mid^2  ( \rho /  \bar{\rho}) \right)
\end{eqnarray}
and unitarity requires
\be
\delta' = \left( \sum_f A^*_{K'_S \to f} A_{K'_L \to f} \right)
/ (\Gamma' + i \Delta m')
\ee
with $\Gamma' + i \Delta m' = i ( \lambda'_L - \lambda^{'*}_S)$ and the
sum extending both to weak decays (as in $\delta$, eq (7)) and to strong
interaction collisions.
\par
To clarify the relationships among these three basis,
two limiting cases of absorbers are worth considering. For a very soft,
low-density absorber ($\nu, \rho \to 0$, $\bar{\rho} \to 1$) one
obviously has
$\mid K'_{S,L}\rangle \to \mid K_{S,L}\rangle$, as seen from
eqs (2) and (8). For a drastic, high-density ("ideal", in ref.
\cite{yogi}) absorber, strong interactions dominate over weak decays
(i.e., $\nu, \ \mid \rho \mid, \ \mid \bar{\rho} \mid \to \infty$),
and one now has $\mid K'_S\rangle \to  \mid \bar{K^0}\rangle$ and
$\mid K'_L\rangle \to \mid K^0\rangle$,
as expected and immediately seen from eqs (8).
\vskip 1cm
{\bf 3. Computing $dP_c/dt$}
\par
Entangled kaon pairs coming from a $\phi$-decay at $t =0$ are described
by eq (1) or, equivalently, by
\be
\mid \Phi (t=0) \rangle = \frac{1}{\sqrt2} \frac{1 + \mid r \mid^2}{2r}
\left[ \mid K_L \rangle \mid K_S \rangle - \mid K_S \rangle \mid K_L \rangle
\right],
\ee
with future time evolution in free space given by eqs (5). The
decay rate into a given c-channel, at time $t$, on
the left beam of the previously described
configuration, is the sum of two contributions:
$dP_{c,\bar{\rho}=1}(t,T)/dt$ and $dP_{c,\bar{\rho} \not= 1} (t,T)/dt$.
The former is the probability rate for c-channel decay (at $t$),
with an accompanying weak decay taking place along
the right beam  between time $0$ and $T$; the second
contribution corresponds to the c-channel decay rate with
accompanying absorption after $T$ (decay or scattering inside the long
absorber). Both separated contributions are
expected to depend on $t$ ( when
c-decay occurs) and $T$ (when the absorber is reached).
Indeed, the use of standard formulae leads unambiguosly to
\begin{eqnarray}
\lefteqn{dP_{c,\bar{\rho}=1}(t,T)/dt = } \nonumber \\
& & \frac{1}{8}
\left( \frac{1}{\mid r \mid}
+ \mid r \mid \right)^2
\Biggl[ \Gamma^c_L (1 - e^{-\Gamma_S T}) e^{-\Gamma_L t}
     + \Gamma^c_S (1 - e^{-\Gamma_L T}) e^{-\Gamma_S t} + \nonumber \\
& & 2 \sqrt{\Gamma^c_L \Gamma^c_S} e^{-\Gamma t}
\frac{1 - \mid r \mid ^2}{1 + \mid r \mid ^2} \left( e^{-\Gamma T}
\cos (\varphi + \Delta m (T-t)) - \cos (\varphi - \Delta m \ t) \right)
\Biggr]
\end{eqnarray}
where $\Gamma^c_{S,L}$ are the c-channel $K_{S,L}$ partial widths and
$\varphi$ is the phase of the ratio of the corresponding amplitudes,
$A_{K_L \to c} / A_{K_S \to c} \ ( \simeq \epsilon \  = \
e^{i \varphi} \mid \epsilon
\mid, \varphi \simeq \pi/4 $, for the typical channels $c = \pi^+ \pi^-
, \ \pi^0 \pi^0$, neglecting $\epsilon'$-effects).
\par
The computation of the second contribution is somewhat more subtle. One
has to use the $K_{S,L}$, free-space basis and
require no decay along the right beam before the edge of the absorber
is reached at time $T$. At this point, the free-space propagating
$K_{S,L}$ states have to be written in terms of
the $K'_{S,L}$ basis, being the appropriate one to study
inside matter propagation. Finally, $K'_{S,L}$ absorption takes places
between $T$ and infinity as described by eq (12). All this implies
\begin{eqnarray}
\lefteqn{dP_{c,\bar{\rho} \not= 1} (t,T)/dt =
\frac{1}{8} \left( \frac{1}{\mid r \mid^2} + 1 \right)
\frac{1}{1 + \bar{\rho}^2} \int_0^{\infty} dt' \sum_f } \\
& & \left| A_{K_L \to c} e^{-i \lambda_L t} e^{-i \lambda_S T} \times \right.
\nonumber \\
& &\left[ (1 + \mid r \bar{\rho} \mid^2 )^{1/2} (1 + \bar{\rho})
A_{K'_S \to f}  e^{-i \lambda'_S t'}
+ (1 + \mid r /\bar{\rho} \mid^2 )^{1/2} \bar{\rho} ( \bar{\rho} -1)
A_{K'_L \to f} e^{-i \lambda'_L t'} \right] \nonumber  \\
& &- A_{K_S \to c} e^{-i \lambda_S t} e^{-i \lambda_L T} \times
\nonumber \\
& &\left. \left[ (1 + \mid r \bar{\rho} \mid^2 )^{1/2} (1 - \bar{\rho})
A_{K'_S \to f}  e^{-i \lambda'_S t'}
+ (1 + \mid r /\bar{\rho} \mid^2 )^{1/2} \bar{\rho} ( \bar{\rho} +1)
A_{K'_L \to f} e^{-i \lambda'_L t'}  \right] \right|^2 , \nonumber
\end{eqnarray}
where the origin of integration over $t'$ has been shifted from
$T \to 0 \  (t' \to t' -T)$ to simplify the notation. Integrating up to
infinity eliminates any dependence on $\bar{\rho}$, i.e., on the kind
of absorber used, and finally leads to
\begin{eqnarray}
\lefteqn{dP_{c,\bar{\rho} \neq 1}(t,T)/dt = } \nonumber \\
& &  \frac{1}{8}  \left( \frac{1}{\mid r \mid}
+ \mid r \mid \right)^2 \Biggl[
 \Gamma^c_L  e^{-\Gamma_L t} e^{-\Gamma_S T}
     + \Gamma^c_S  e^{-\Gamma_S t} e^{-\Gamma_L T} \nonumber \\
&-& 2 \sqrt{\Gamma^c_L \Gamma^c_S} e^{-\Gamma t} e^{-\Gamma T}
\frac{1 - \mid r \mid ^2}{1 + \mid r \mid ^2}
\cos (\varphi + \Delta m (T-t)) \Biggr] .
\end{eqnarray}
\par
As expected, the two separated contributions (16) and (18)
depend on both $t$ and $T$. However, their sum, which corresponds to
the the reading of the c-channel decay detector, turns out to be
\be
\frac{dP_c (t,T)}{dt} =
\frac{1}{8} \left( \frac{1}{\mid r \mid}
+ \mid r \mid \right)^2
\left[ \Gamma^c_L  e^{-\Gamma_L t}
     + \Gamma^c_S  e^{-\Gamma_S t}
- 2 \sqrt{\Gamma^c_L \Gamma^c_S} e^{-\Gamma t}
\delta \cos (\varphi - \Delta m \ t) \right]
\ee
with no $T$-dependence.
\vskip 1cm
{\bf 4. Discussion}
\par
Unfortunately our results disagree with the much more interesting ones
obtained by Srivastava et al. \cite{yogi} in essentially the same
context. Indeed, our $T$-independence in eq (19) will be preserved if
additional $t$-integrations between $t_i$ and $t_f$ with
$T$-independent cuts are performed as in ref. \cite{yogi}. To
understand the origin of the discrepancy, we particularize eq (17) to
the "ideal" absorber case, i.e., to the case of immediate strong
absorption at $T$ (no $t'$-integration in eq (17) is thus required)
with identification of a $K^0$ $versus$ a $\bar{K^0}$ (the only two
terms surviving in the summation). Eq (17) now becomes
\begin{eqnarray}
\lefteqn{dP_{c, ideal} (t,T)/dt =
\frac{1}{8} \left( \frac{1}{\mid r \mid } + \mid r \mid  \right)^2 } \\
& &\Biggl[ \left| A_{K_L \to c} e^{-i \lambda_L t} e^{-i \lambda_S T}
\langle K^0 \mid K_S \rangle \
- \   A_{K_S \to c} e^{-i \lambda_S t} e^{-i \lambda_L T}
\langle K^0 \mid K_L \rangle  \right|^2  \nonumber \\
&+& \left| A_{K_L \to c} e^{-i \lambda_L t} e^{-i \lambda_S T}
\langle \bar{K^0} \mid K_S \rangle \
- \  A_{K_S \to c} e^{-i \lambda_S t} e^{-i \lambda_L T}
\langle \bar{K^0} \mid K_L \rangle
\right|^2 \Biggr], \nonumber
\end{eqnarray}
leading again to the r.h.s. of eq (18) if eqs (2) are used to obtain
the four $\langle K^0, \bar{K^0} \mid K_{S,L} \rangle$ projections with their
corresponding phases. If the sign difference in the $\pm r$ term
in eq (2) is (injustifiably, in our opinion)
ignored we reproduce all the results derived in ref.
\cite{yogi}, where the use of projection operators was avoided.
\par
Although in our discussion we have always assumed $t < T$,
our final $T$-independent result (19) obviously holds for $T \leq t$ as
well, $i.e.$, T-reversal non-invariance is of no help here.
In this sense, we fully agree with the general theorems or
conclusions proposed
in refs. \cite{several}, \cite{ghirardi}, namely, that the results
(at the statistical or observable level) obtained in one of the beams
in an EPR-like experiment cannot be modified by acting along the
other beam. This general conclusion is usually deduced from the basic
formalism of Quantum Mechanics which is not entirely free from
subtleties and controversies, as exemplified in refs. \cite{datta}-
\cite{ghirardi} and \cite{selleri}. We feel that the possible
significance of our paper lies in the fact that -- as in the
interesting analysis by Srivastava and Widom, that triggered our
present reconsiderations -- only standard formulae and an explicit and
widely accepted procedure have been used.

\vskip 1.5cm

{\bf Acknowledgements}
\par
Thanks are due to Y. Srivastava and A. Widom for very interesting
discussions.
This work has been supported in part by the  Human Capital and
Mobility Programme, EEC Contract \# CHRX-CT920026 and, by DGICYT,
AEN93/0520.


\newpage


\begin{thebibliography}{99}

\bibitem{dafne} The DA$\Phi$NE Physics Handbook, eds. L. Maiani,
G. Pancheri and N. Paver, Laboratori Nazionali di
Frascati, June 1992.
\bibitem{datta} A. Datta, D. Home and A. Raychaudhuri,
Phys. Lett. $\underline{A123}$ (1987) 4.
\bibitem{datta2} A. Datta, D. Home and A. Raychaudhuri,
Phys. Lett. $\underline{A130}$ (1988) 187.
\bibitem{several} M. J. W. Hall,
Phys. Lett. $\underline{A125}$ (1987) 89; \\ G. Lindblad,
Phys. Lett. $\underline{A126}$ (1987) 71; \\ E. Squires and D. Siegwart,
Phys. Lett. $\underline{A126}$ (1987) 73; \\ J. Finkelstein and H. P.
Stapp, Phys. Lett. $\underline{A126}$ (1987) 159.
\bibitem{ghirardi} G. C. Ghirardi et al.,
Eurohys. Lett. $\underline{6}$ (1988) 95 and references therein.
\bibitem{yogi} Y. Srivastava and A. Widom,
Phys. Lett. $\underline{B314}$ (1993) 315.
\bibitem{etp} A. Einstein, R. C. Tolman and B. Podolsky,
Phys. Rev. $\underline{37}$ (1931) 780; \\
A. Einstein, B. Podolsky and N. Rosen,
Phys. Rev. $\underline{47}$ (1935) 777.
\bibitem{kabir} P. K. Kabir, "The CP Puzzle", Academic Press, N.Y.
(1968).
\bibitem{selleri} F. Selleri, "Quantum Paradoxes and Physical Reality",
Kluwer Academic Publishers (1990).

\end{thebibliography}
\end{document}